# Tunable viscosity modification with diluted particles: When particles decrease the viscosity of complex fluids.

Manuchar Gvaramia[1,2], Gaetano Mangiapia[1], Vitaliy Pipich[1], Marie-Sousai Appavou[1], Gerhard Gompper[3], Sebastian Jaksch[1], Olaf Holderer[1], Marina D. Rukhadze[4], Henrich Frielinghaus[1,*]

[1]Jülich Centre for Neutron Science at MLZ, Forschungszentrum Jülich GmbH, Lichtenbergstrasse 1, 85747 Garching, Ger-many
[2]Ilia Vekua Sukhumi Institute of Physics and Technologies, 7 Mindeli Str., 0186, Tbilisi, Georgia
[3]Institute of Complex Systems 2, Forschungszentrum Jülich GmbH, 52425 Jülich, Germany
[4]Department of Chemistry, Ivane Javakhishvili Tbilisi State University, Chavchavadze Ave. 3, Tbilisi, 380028, Georgia

**ABSTRACT:** While spherical particles are the most studied viscosity modifiers, they are well known only to increase viscosities, in particular at low concentrations. Extended studies and theories on non-spherical particles find a more complicated behavior, but still a steady increase. Involving platelets in combination with complex fluids displays an even more complex scenario that we analyze experimentally and theoretically as a function of platelet diameter, to find the underlying concepts. Using a broad toolbox of different techniques we were able to decrease the viscosity of crude oils although solid particles were added. This apparent contradiction could lead to a wider range of applications.

**Introduction.** The demand of viscosity modifications by solid particles has many applications in the field of food science [1] and oil recovery [2]. The simplest theoretical description was found by Einstein for diluted spherical particles [3]. At that point, the predicted changes of the viscosity were small, especially for large dilutions in the concentration range of 1%vol. However, applications usually manifold either at large concentrations (as in the case of chocolate [4]) or in asymmetric particles, such as clay, which has many uses [5]. Therefore, the theoretical concepts became more complicated [6,7], and the large number of possibilities opens up a wide range of inter-pretations.

The medium in nearly all these cases has been assumed to be a simple liquid. However, the response of complex fluids [8,9] to particles [10,11] can be more complicated [12,13], and, therefore, back coupling might be possible. In our past studies we have studied the lubrication effect, which describes the lamellar ordering [14] of a complex fluid, i.e a microemulsion (µE) next to a hydrophilic wall. This ordering allows the domains for facilitated sliding along the surface [13]. In neutron scattering studies we have found that the typical relaxation times in these lamellae are also faster than in the bulk. So for platelet particles, a facilitated flow behavior (lower viscosity) might be expected [12]. The change of platelet diameters revealed that the perfection of lamellar modes is increasing with increasing platelet diameter. Consequently, large platelets would serve the aim of reduced viscosities better than small ones.

Until now, the studied liquids consisted of components with highly similar viscosities. Therefore, the rearrangement of domain structures resulted in similar viscosities, and the highly desired case of viscosity reduction was not found.

We recall the theoretical concepts that are known, show an interesting rheology experiment, and prove some structural prerequisites (see Figure 1). Then, we explain the rheological findings. Finally, we show the example of cooled crude oils, where astonishingly a viscosity decrease is found.

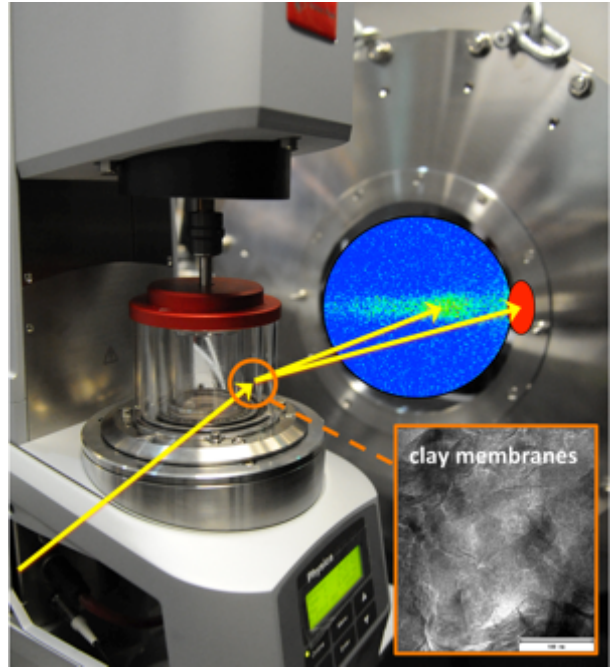

**Figure 1:** A sketch of the combination of SANS with rheology at the instrument KWS1 (and KWS3). A TEM image inlay indicates the clay membranes that we observe in solution.

**Theory of particles.** The viscosity modification of simple fluids by spherical particles is well studied and well understood. While at low concentrations the simple law of Albert Einstein [3] is the essential result, high concentrations are more interesting for industrial applications. One simple formula is for example [6]:

$$\eta_{\text{rel}} = \frac{\eta}{\eta_0} = \left[1 - \frac{\phi}{1-c\phi}\right]^{-[\eta]} \tag{1}$$

On the left side we find the relative viscosity with respect to the pure simple fluid ($\eta_0$). The term in the brackets depends on the particle concentration $\phi$ and the crowding factor $c = \phi_c^{-1} - 1$, which is dependent on the maximum concentration of the particles, i.e. $\phi_c \approx \phi_{\max} = 0.74$ (or 0.64 [7]) for monodisperse spherical particles. The exponent $[\eta]$ is called intrinsic viscosity and takes the value 5/2 for spherical particles. Thus, in the limit of smallest concentrations, the Einstein formula [3] is obtained: $\eta_{\text{rel}} = 1 + 5/2\ \phi$.

The essential result of equation 1 is that (a) the viscosity always increases with increasing particle concentration, and that (b) the relative viscosity is weakly dependent at small particle concentrations.

A more detailed analysis of the intrinsic viscosity $[\eta]$ for anisotropic particles introduces the aspect ratio $A_f$ as the essential parameter. The value $A_f$ takes larger values than 1 for prolate ellipsoids. Fibers are obtained for $A_f \gg 1$. Conversely, our interest focuses on oblate particles ($A_f < 1$) that take the limit of $A_f \ll 1$ in the case of clay particles. The intrinsic viscosity takes the limit [7]

$$[\eta] = \frac{1012}{1497} \cdot A_f^{-1} \approx 0.68 \cdot A_f^{-1} \tag{2}$$

for very oblate particles. In this limit, the packing gets rather dense, and the packing parameter c ≈ 0 nearly vanishes. Thus the expected behavior results in:

$$\eta_{\text{rel}} = [1 - \phi]^{-0.68 \cdot A_f^{-1}} = [1 - \phi]^{-0.68 \cdot D/t} \tag{3}$$

Here the platelet diameter is *D* and its thickness is *t*. At particle concentrations of ϕ = 1% and aspect ratios of $A_f$ = 0.01 the predicted viscosity increase is in the range of 2.

This result is based on the argument of very high packing capabilities of platelets. Arguing on the basis of percolation [7], the critical concentration ϕc is considerably lower than unity. Summarizing and applying their formulas for platelets, the following critical concentration is obtained:

$$\phi_c = \frac{9.875 \cdot 1.658}{7.742}(-2 \ln A_f)A_f = -4.2 A_f \ln A_f \tag{4}$$

Taking a typical aspect ratio of $A_f$ = 0.01 results in still relatively high ϕc ≈ 0.2 that have no major implications with respect to the first assumption of very high packing. However, from experiments we know that gelation of clay particles takes place in the range of 1 to 2% volume fraction [5]. This discrepancy is due to the charges on the platelets, so the pure geometric considerations break down for real clay platelets.

Crossing over to the gel state, the system becomes viscoelastic with the proportionality of the complex shear modulus being:

$$G \sim (\phi - \phi_c)^\gamma \tag{5}$$

with the exponent γ being in the range of 0.8 to 2. Since the already defined concentration ϕc becomes rather constant with changing platelet diameter, equation 5 would predict a rather constant modulus. Again, the pure geometric consideration breaks down, and more advanced theories are needed. Here we point out that the scaling of the viscosity η and the shear modulus *G* are treated in this manuscript on a similar basis independent of the actual phase that is considered (i.e. fluid and gel phase). The argument is that the microscopic frequencies of the motions affect both magnitudes in the same manner, and the scaling with changing platelet diameter should be the same. Further, from the rheology of our systems we can confirm that we reached the gel phase where the complex viscosity |η*| is proportional to the reciprocal frequency $\omega^{-1}$ over a wide range of time scales in agreement with Reference 15. Thus, our selected complex viscosity is rather specific for the gel modes, while the steady shear viscosity is mostly influenced by the slowest modes. In our case, these two approaches should agree.

**Theory of lamellar membranes.** Another theory was developed for lamellar systems of membranes in a fluid medium [16]. Since in microemulsions, the thermodynamics is dominated by the behavior of the surfactant membranes only, the same concepts hold for lamellar micro-emulsions [16,17]. Such lamellar microemulsions form in the vicinity of hydrophilic planar surfaces [14] and especially in the presence of clay particles [12]. The viscosity ratio between two lamellar systems results from [16]:

$$\eta_{\text{rel}} = \frac{\eta_1}{\eta_2} = \frac{E_2 \Lambda_2}{E_1 \Lambda_1} \tag{6}$$

With the energetic term $E_i$ and the Oseen tensor $\Lambda_i$. The energetic term reads:

$$E_i = \kappa_i k^4 + \frac{1}{\kappa_i d_{E,i}^4} \tag{7}$$

with the membrane bending rigidity $\kappa_i$ normalized to the thermal energy $k_B T$, the wavevector $k$ of the critical mode (connected to the platelet diameter), and the membrane distance $d_{E,i}$ that determines an offset energy. The Oseen tensor is given by:

$$\Lambda_i = \frac{1}{4k\eta_i} \cdot \left(1 - \frac{1 + 2d_i k + 2(d_i k)^2 (1 + 2L_i k)}{(1 + 4L_i k)\exp(2d_i k)}\right) \tag{8}$$

with the viscosity $\eta_i$ of the fluid around the membranes in system *i*, and the perforation dimension $L_i$. On tiniest length scales ($k \rightarrow \infty$) the viscosity of the medium is the only contribution. The bracket describes a peaked function with the maximum around k = 0.9/$d_i$, which is connected to the repeat distance di. Introducing perforations keeps the maximum lower and moves it to larger *k*. However, this maximum describes a facilitated flow along the oriented lamellae.

**Materials.** Clay particles were obtained from Süd Chemie (Rockwood), now belonging to the BYK chemicals group. We received four types of clay: Laponite RD that is reported to have a diameter of 25-30nm, Laponite OG that is reported to have a diameter of 80nm, Laponite WXFN that is reported to have a diameter of 140nm, and Montmorillonite EXM 757 had a diameter of ca. 500nm. The Laponite materials were used without further cleaning. The Montmorillonite material was dispersed in deionized water by sonication over night, centrifuged, and the liquid portion dried using a rotor vap and a vacuum oven. The particle characteristics are summarized in Table 1.

**Table 1:** Clay particles from Südchemie / Rockwood

| Name | Abbreviation | Diameter [nm] | Thickness [nm] |
|---|---|---|---|
| Laponite RD | LRD30 | 30 | 1 |
| Laponite OG | LOG80 | 80 | 1 |
| Laponite WXFN | WXFN140 | 140 | 1 |
| Montmorillonite EXM 757 | MMT500 | 500 | 1 |

N-decane and NaCl were obtained from Sigma Aldrich. The surfactant $C_{10}E_4$ was obtained from Bachem, Weil am Rhein, Germany. Deuterated n-decane and heavy water were obtained from Armar chemicals, Döttingen, Switzerland. All these chemicals were used without further purification. Deionized water was obtained from the Purelab Ultra filter from ELGA at 18.2MΩcm.

The crude oils from Pennsylvania and Colorado (Denver basin) were obtained from ONTA Inc., Toronto, Canada. They were used as received.

1%vol clay dispersions were obtained by mixing the clay with deionized water and 1%wt NaCl, and sonication over night. 1%vol clay and 1%wt NaCl were also dispersed in a microemulsion with a stock of 17%vol $C_{10}E_4$, 41.5%vol water, and 41.5%vol n-decane. For neutron measurements we used heavy water for the aqueous clay dispersion, and the bulk contrast in the microemulsion samples. Film contrast with deuterated n-decane and heavy water was investigated as well – but no useful results emerged.

The crude oils were sonicated with the Montmorillonite clay over three days. Only then was a viscosity de-crease obtained for all samples. By combusting the liquid phase, a MMT500 solubility of 1.1 to 1.2% and 0.6 to 0.7%wt was determined for the Pennsylvania and Colorado crude oils, respectively (see also Supporting Information).

**Experimental Methods**. Rheology experiments were conducted on a MCR302 and MCR501 from Anton Paar [18,19]. The MCR302 could theoretically detect ten times lower viscosities than the MCR501. The storage and loss moduli G' and G'' were measured using a Couette cell that could also be mounted on a SANS instrument with an outer diameter of 50mm and an inner diameter of 49mm and a height of 60mm. The large volume and surface supported much more sensitive measurements that were needed in the case of the very fluid samples. All measurements took place in the linear regime with a strain of $\gamma = 1\%$ (strain controlled); they were conducted in a frequency sweep from higher to lower frequencies (630 to $10^{-3}$ rad/s) without further pretreatment. The complex viscosity was calculated on the basis of $\eta^* = (G'+iG'')/(i\omega)$. One very high viscosity sample lead to the destruction of the system, and, thus, the crude oils were characterized in a cone-plate geometry with 5cm diameter and 1° inclination angle. The strain amplitude was $\gamma = 1\%$. All crude oils were measured at constant shear frequency $\omega=10$rad/s from higher to lower temper-atures. The rheological SANS measurements were con-ducted under steady shear with given shear rates of 0.1, 1, 10, and 100s$^{-1}$.

Small Angle Neutron Scattering (SANS) measurements were conducted on KWS1 [20,21] and KWS3 [22,23] at the FRM-II of the Heinz Maier Leibnitz Zentrum in Garching. The rheometer was installed for mainly tangential observations, where the alignment of the clay particles could be studied, but radial measurements were conducted as well. The neutron wavelength was 0.5nm for KWS1 and 1.28nm for KWS3. KWS1 is a classical pin-hole SANS for a dynamic Q-range from ca. $10^{-2}$ to 5nm$^{-1}$. KWS3 uses a focusing mirror that focuses on the detector to reach lower Q from $5\times10^{-4}$ to $5\times10^{-2}$ nm$^{-1}$. On KWS3 we went off-center with the detector in order to reduce the background of the primary beam. Furthermore, we had to apply a prism correction [24] due to the tangential geometry with the large wavelengths. All of this resulted in a Q-range of ca. $10^{-2}$ to $5\times10^{-2}$nm$^{-1}$. Absolute calibration was done as good as possible, since the sample thickness in the tangential geometry is not known to a high precision.

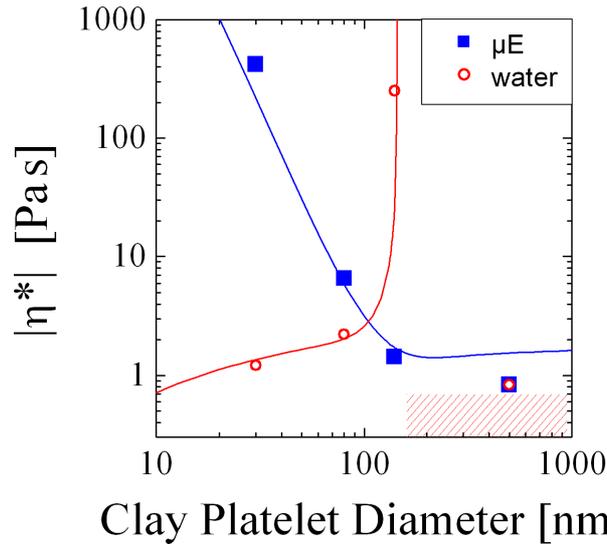

**Figure 2:** The modulus of the complex viscosity of clay dispersions in water (red) and in a microemulsion (blue, µE) as a function of the platelet diameter at 1%vol particle concentration.

**Results and Discussion.** Extensive oscillatory rheological measurements were conducted to characterize all fluids. For most of the samples, we observed a gel-like behavior with a viscosity $|\eta^*|$ scaling with the reciprocal frequency $\omega^{-1}$ [15] within $\omega$ = $10^{-3}$ to 50 rad/s in the linear regime ($\gamma$=1%, see Supporting Information, also for discussion of instabilities). The selected viscosities at 1 rad/s are displayed in Figure 2. For the aqueous systems, we observe an increase of the viscosity with increasing platelet diameter, while the opposite is observed in microemulsions. The solid lines are obtained from discussions below and at this stage are a guide for the eye.

At low particle concentrations of 1%vol used here, the aqueous samples start at moderate viscosities for the smaller platelet diameters of 30 and 80nm when compared to the viscosity of water ($10^{-3}$ Pa s). But even in this case, the classical geometrical theories cannot predict a gain of ca. 1000. Astonishingly, the gain for the larger particles of 140nm diameter lies in the range of $3\times10^5$. This looks like a criticality that shall be discussed below. The general trend of increasing viscosities agrees with equation 3, although the numbers do not directly lead to such high viscosity gains as experimentally found. The Montmorillonite with 500nm diameter seems to be off this trend, and must be considered separately (also indicated by the shaded red region).

The decreasing viscosities for the microemulsions can be seen in context with the lubrication effect [13,25]. We have seen that the lamellae are developed better along large platelets. This would facilitate the off-sliding of the domains much more at larger platelet diameters, as observed by rheology. So the local ordering of the microemulsion as a medium serves for facilitated flow along the oriented platelets along the macroscopic flow. The application of the theory (as shown by the solid blue lines) will be discussed below.

All in all we have seen two opposite trends of the viscosities depending whether we take a simple or complex fluid as medium for the platelets. This new insight could only be revealed by testing different platelet diameters.

To access the microscopic parameters that we need for modeling the viscosity obtained from the rheology experiment, SANS experiments have been conducted using steady shear. The principles have already been explained elsewhere [26,27], and the explicit ordering of clay platelets can be found in the literature [28,29,30]. The needed length scale range is estimated through $\ell = t/\phi$, and $Q_{min} = \ell^{-1}$, which lies in the range of $10^{-2}$ nm$^{-1}$ (for a platelet thickness $t$ = 1nm, and a concentration $\phi$ = 0.01). The results from the two instruments KWS1 and KWS3 are summarized in Figure 3. While KWS1 already probes the classical SANS Q-range down to ca. $10^{-2}$nm$^{-1}$, KWS3 surely extended the Q-range further down. However, only after the experiment did the refractive effects of the tangential geometry (and longer wavelengths) become clear, and were corrected afterwards.

The aqueous dispersions show a clear Bragg peak at $Q_{peak} \approx 0.02$nm$^{-1}$ with the sheet normal direction (and repeat unit) being oriented along the shear gradient. A weak second peak is distinguished at $Q_{peak} \approx 0.03$nm$^{-1}$. The anisotropy clearly indicates a preferential lamellar order of sheets. The latter weaker peak already appears for the sample without shear. So if there is a population with a shorter repeat distance, it is less important, and might form at the walls of the Couette cells. For the LRD30 we also managed to have higher order Bragg peaks of the orders 3, 4, and 5 (The second order might be missing due to the strong unwanted background that decays with increasing Q. One origin of the parasitic scattering results from the Couette cell wall roughness at grazing incidence angles.). The slight mismatch in Q might result from the incomplete prism correction of the tangential geometry that might yield a slightly lower first order Bragg $Q_{max}$. In any case, the repeat distance of clay sheets is about 300nm, which is much larger than initially expected from isolated platelets (ca. 100nm). Thus, the platelets form tactoid structures with approximately 3 platelets overlapping in the normal direction. This principal result was already found elsewhere [26] with slightly different numbers. The regular liquid crystalline order develops only when shear is applied.

The weak scattering signals are still visible due to fortunate side effects: In case of the aqueous systems, the enlarged thickness of the tactoids increases the scattering intensity. In the case of the microemulsion systems, the preferential neighborhood of water (D$_2$O) next to the platelets doubles the effective thickness of deuterated material. Thus the series of layers effectively appears to be like …HDHDcDHDH… (c for clay of 1nm thickness, and the oil(H) and water(D) domains of ca. 10nm). The same configuration was not visible in film contrast where the oil and water were deuterated, because this arrangement effectively appears as …DfDfDcDfDfD… (now c for clay, D for either oil or water, and f for surfactant film of ca. 1nm). The local lamellar ordering of domains is still superimposed by a large fraction of bicontinuous microemulsion without preferred orientation.

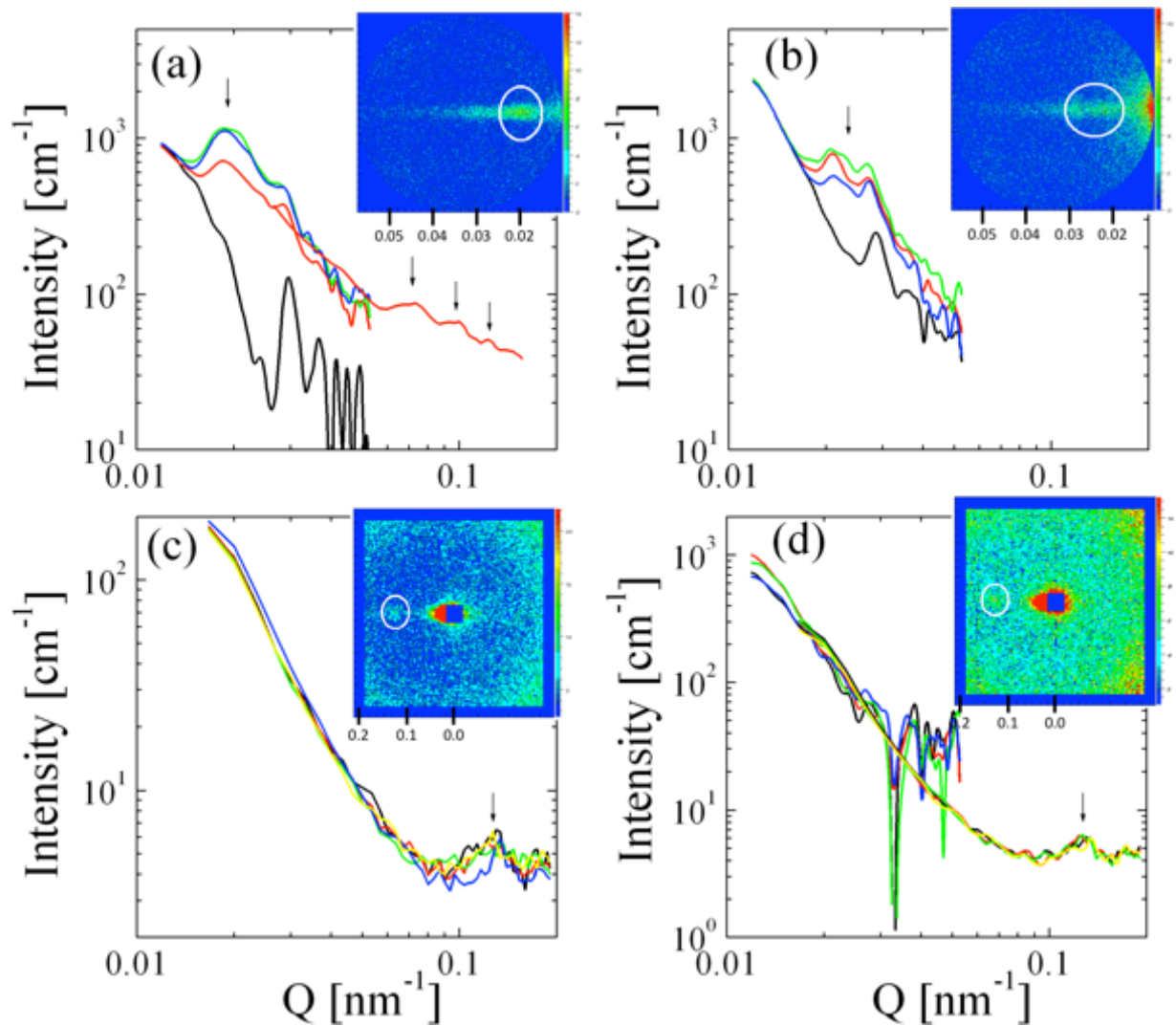

**Figure 3:** SANS curves from the different clay suspensions, intensity as a function of the scattering vector $Q$ along the shear gradient: (a) LRD30 in $D_2O$ for no shear (black), 0.1s$^{-1}$ (red), 1s$^{-1}$ (green), and 10s$^{-1}$ (blue). (b) MMT500 in $D_2O$ for no shear (black), 0.1s$^{-1}$ (red), 1s$^{-1}$ (green), and 10s$^{-1}$ (blue). (c) LRD30 in bulk contrast microemulsion ($D_2O$ only) for no shear (black), 1s$^{-1}$ (red), 10s$^{-1}$ (green), 100s$^{-1}$ (blue), and 0.1s$^{-1}$(yellow). (d) MMT500 in bulk contrast microemulsion ($D_2O$ only) for no shear (black), 1s$^{-1}$ (red), 10s$^{-1}$ (green), 100s$^{-1}$ (blue), and 0.1s$^{-1}$(yellow). Inlays in (a,b)/(c, d) show 2d detector images from SANS at 10m on KWS3/ at 20m on KWS1 and indicate the Bragg peak of an oriented sample. Arrows in general indicate Bragg peaks of clay sheets with the normal direction along the shear gradient. The vertical detector axis corresponds to the rotational axis rot($\vec{v}$).

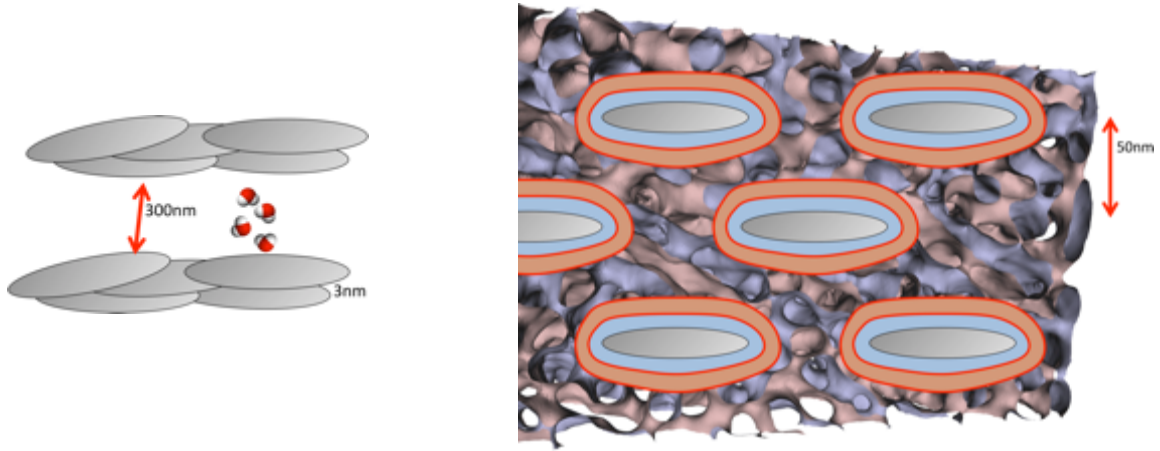

**Figure 4:** The interpretation of the scattering experiments in terms of clay platelet structure in the two fluids: simple (left) with tactoids and complex (right) with checkerboard structure of isolated platelets (the red and blue colors indicate oil and water domains).

Now that the structure of the clay particles is determined, we can interpret the rheological experiments. The platelets form tactoids in the aqueous systems that resemble a membrane of 3nm thickness and 300nm repeat distance. The two structural levels are (1) the tactoid structure embedded in (2) water. The platelets in the microemulsion systems form highly perforated membranes of 1nm thickness and 50nm repeat distance. The two structural levels are (1) clay membranes embedded in (2) a microemulsion. For both levels we will apply the (modified) lamellar descriptions of equations 6-8. The used parameters are listed in Table 2. The agreement with the measurements is quite good as seen in Figure 2. The essential parameter is the cutoff wavelength $k = 2\pi/D$ that is determined by the platelet diameter D.

**Table 2:** The parameters used to describe the rheological data.

| Parameter | Aqueous system | Microemulsion system |
|---|---|---|
| prefactor $\eta_2 = \eta_{liq.}$ | 0.06 Pa s | $4.5 \times 10^5$ Pa s |
| k | $2\pi/D$ | $2\pi/D$ |
| $\kappa_1$ | $\ln(D/3nm)/(4\pi)$ [$k_BT$] | $\ln(D/1nm)/(4\pi)$ [$k_BT$] |
| $d_{E,1}$ | 1nm | 1nm |
| $d_1$ | 300nm | 50nm |
| $L_1$ | $(\sqrt{2}-1)\times D$ | D |
| screening k | $2\pi/(5\times D)$ | normal $2\pi/D$ |
| $\kappa_2$ | 1 [$k_BT$] | 1 [$k_BT$] |
| $d_{E,2}$ | 0.3nm | 20nm |
| $d_2$ | membrane modes | 10nm |
| $L_2$ | omitted | D |

The prefactor $\eta_2$ of the aqueous system ideally takes the medium viscosity $\eta_{liq.} = 10^{-3}$Pa s while we find a 60 times larger value that is due to the gel state of the system. The large prefactor of the microemulsion systems relates to $d_{E,2}$, the microemulsion repeat distance which becomes important at large platelet diameters. So we argue, that the prefactor should be extended by an explicit factor of $(d_{E,2}(microemulsion)/d_{E,2}(pure\ water))^4 = (20/0.3)^4$, the numbers of which are discussed below, in concert with the microemulsion viscosity of 0.01Pa s, resulting in ca. $2\times10^5$ Pa s. The bending rigidity $\kappa_1$ of the clay membrane is determined over the persistence length $\lambda = D = t\exp(4\pi\kappa_1)$ [26], that we identify with the platelet diameter, and the membrane thickness $t$. The offset energy in the energetic term $E_1$ is connected to the thickness $d_{E,1}$ that we identify with the single platelet thickness of 1nm. In contrast, in the Oseen tensor the fluctuations relate to the repeat distance of the clay membranes that we determined by SANS. The holes in the clay membranes are either connected to omitted platelets in the vicinity of 4 neighboring platelets, i.e. $L_1 = (\sqrt{2}-1)D$, or to the checkerboard voids, i.e. $L_1 = D$. The screening of the modes was taken as originally given in the case of the checkerboard membranes, but 5 times extended for the tactoids, which develop extended surfaces.

The reference system water was simply modeled by the offset energy $d_{E,2}^4$ that we connect to the molecule size of ca. 0.3nm. The reference microemulsion system is described by a bending rigidity 1 [31], the offset energy related dimension of the full repeat distance $d_{E,2}$ = 20nm, and the membrane modes of each surfactant membrane. The perforations that are connected to the checkerboard structure are kept in this circumstance.

The main result of this modeling is the fact that the clay sheets form membranes (either tactoids or checkerboard sheets) that determine the rheological response of the overall system. The reference system is either a simple liquid or a complex fluid. This second contribution is also essential for the general rheological response. In the case of simple fluids, the viscosity increases with the platelet dimension. In the case of complex fluids, the viscosity decreases with the platelet dimension.

The lowest viscosities of the aqueous system are only amplified by a factor of ca. 100 at smallest platelet diameters (30 and 80nm), while the same happens for the microemulsion system at 140 and 500nm platelet diameter. The viscosity of the simple fluid is heavily increased at platelet dimensions of 140nm, where the membrane becomes "stronger" (refer to effective $\kappa_1$). The viscosity of complex fluids is heavily increased at smaller platelet dimensions, where the platelets cut off the modes at rather small dimensions. In either case this amplification of viscosities is tremendous, and thus interesting for applications. This effect is by far not covered by the simple geometrical considerations of isolated particles, because it is a collective phenomenon in either case. It is not to be confused with classical boundary layer conditions [32].

The breakdown of the sharp viscosity increase at around 140nm towards larger platelet dimensions is seen in context with the MMT500 platelets. Here, the viscosity takes the lowest values that we obtained for clay systems. The breakdown is connected to the modified screening. If the correlations along the membrane considerably exceed the repeat distance of the tactoids, the guidance of the fluid by the tactoids seems to overwhelm everything. In this sense the system becomes similar to complex fluids where the membranes with large correlation lengths simply guide the fluid.

From crude oils we know that two different effects cause higher viscosities at cooler temperatures compete [33,34] (see also supporting information). The flocculation happens between ca. 100°C and 0°C and results from pi-stacking of aromatic fractions. This fraction remains at intermediate viscosities. The wax crystallization happens below ca. 0°C and causes real solid fractions. Apart from that, there are shorter aliphatic molecules, which stay liquid in our experimental window. The crude oils do not tend to phase separation, which might result from mediating molecules with aromatic and aliphatic properties – similar to surfactants in microemulsions. In the supporting materials we try to argue that the higher amounts of mediating molecules support the formation of lamellar phases, which supports the observed effects. Furthermore, one of the aromatic or aliphatic components will preferentially wet the clay particles.

Motivated by this picture, we tried several crude oils that were available on the market, and dispersed MMT500 clay in them. We present two results from crude oils from Pennsylvania and Colorado, the cooling cycles of which we present in Figures 5 and 6. The first oil shows a simple fluid behavior close to room temperature. At room temperature, the pure oil is more fluid than the one with clay. Below ca. 14°C the situation is reversed, and factors of up to 10 describe the reduction of complex viscosity. In the Supporting In-formation the viscoelasticity of the two fluids can be compared on the basis of the loss tangent. Furthermore we discuss the conditions of the oils for these reduced viscosities.

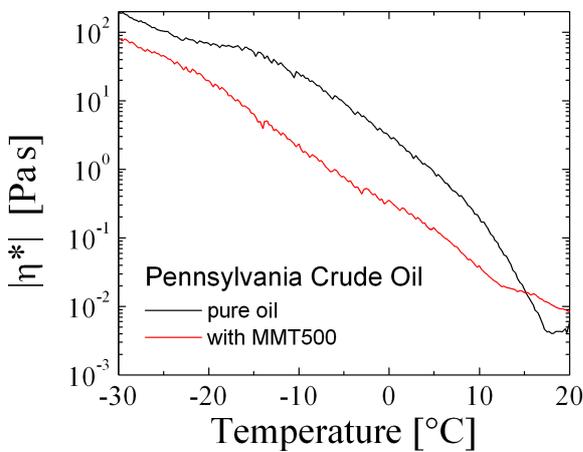

**Figure 5:** The modulus of the complex viscosity of a MMT500 suspension in Pennsylvania crude oil and its reference as a function of temperature.

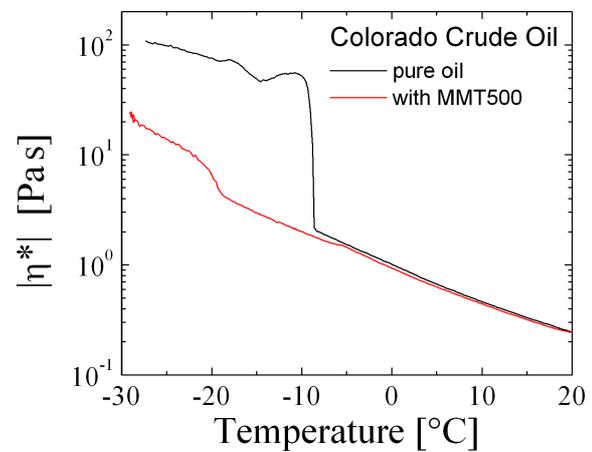

**Figure 6:** The modulus of the complex viscosity of a MMT500 suspension in Colorado crude oil and its reference as a function of temperature.

The Colorado crude oil displays a similar transition at around -8°C, where suddenly the pure oil has a strongly increased complex viscosity. When looking in more details on the loss tangent (Supporting Information), there seem to be more phase transitions that we cannot comment on. However, also here the reduction of the complex viscosity can be a factor of 10.

For the first example with the classical microemulsion as a complex fluid, the viscosities of all components were nearly all the same. In the case of the lowest viscosities with clay particles, the viscosity was elevated by factors of ca. 100 because of the clay. For crude oils, the lamellar alignment of domains along the clay particles serves for strongly reduced viscosities with respect to the original reference system without clay. The reason seems to arise from the very different viscosities of the components. A perfect lamellar system of different components $i$ would have the average viscosity [35,36]:

$$\eta_{\text{liq.}}^{-1} = \sum_i \frac{\phi_i}{\eta_{\text{bare},i}} \qquad (9)$$

while a completely random system would follow the Arrhenius formula [36,37]:

$$\eta_{\text{liq.}} = \alpha \prod_i \eta_{\text{bare},i}^{\phi_i} \qquad (10)$$

where the factor α lies in the range of 1 to 10 for many systems (Supporting Information). The physical interpretation is connected to the heat or mixing [38] $Q_m$, according to

$$\alpha = \exp\left(-\int \frac{Q_m}{k_B T^2} dT\right) \qquad (11)$$

The temperature T is given in absolute units, and $k_B$ is the Boltzmann constant. Practical ranges of α are discussed in the Supporting Information.

The different way of averaging can be understood in the simple example: We have a symmetric 2-phase system ($\phi_i$ = 0.5) with viscosities of $\eta_0$ and $100\eta_0$. The random system has a viscosity of $10\alpha\eta_0$, while the ordered system has a viscosity of $1.98\eta_0$.

We believe that a lower fraction of mediating molecules lead to a more pronounced formation of aromatic and aliphatic domains that then can be aligned by the clay particles (see Supporting Information).

**Conclusions.** As a result, the crude oils show complex fluid behavior with at least two components of very different viscosities. The clay particles induce a lamellar ordering that finally explains the very different viscosities with and without clay due to alignment of the domains. This alignment causes different averaging of the viscosities for the liquid medium around the solid particles (equations 9 and 10). Thus, large clay particles can be used to dramatically decrease the viscosity of complex fluids. Furthermore, microemulsions can be used as ideal model complex fluids that allow for studying the principal mechanisms in more complicated applied systems.

**ACKNOWLEDGMENT**
We thank Helmut Coutelle from the former Rockwood section, Moosburg, for his kind provision of clay samples. We thank Bernhard Mogge from Forschungszentrum Jülich, Patent Department for fruitful discussions. We thank Stefan Grumbein, formerly from Technical University Munich, Garching, Germany, for performing measurements on the more sensitive MCR302. We thank our chem-ical analysis division ZEA-3, in particular Sabine Willbold and Michael Krinninger for chemical analysis support. This analysis was supervised by Jürgen Allgaier.

*Supporting Information:*

**Tunable viscosity modification with diluted particles: When particles decrease the viscosity of complex fluids.**

Manuchar Gvaramia[1,2], Gaetano Mangiapia[1], Vitaliy Pipich[1], Marie-Sousai Appavou[1], Gerhard Gompper[3], Sebastian Jaksch[1], Olaf Holderer[1], Marina D. Rukhadze[4], Henrich Frielinghaus[1,*]

[1]Jülich Centre for Neutron Science at MLZ, Forschungszentrum Jülich GmbH, Lichtenbergstrasse 1, 85747 Garching, Germany
[2]Ilia Vekua Sukhumi Institute of Physics and Technologies, 7 Mindeli Str., 0186, Tbilisi, Georgia
[3]Institute of Complex Systems 2, Forschungszentrum Jülich GmbH, 52425 Jülich, Germany
[4]Department of Chemistry, Ivane Javakhishvili Tbilisi State University, Chavchavadze Ave. 3, Tbilisi, 380028, Georgia

As discussed in the main text and here, the observations and theories relate to the linear regime of rheological responses. Thus, the shear fields are weak ($\gamma=1\%$). When leaving this range, instability and chaos can occur. On the one hand, a viscosity reduction was found for polymer melts [39], and a rather complex behavior of turbulences [40] could be observed for clay systems.

The full frequency sweeps from high to low frequencies of the rheological measurements are shown in Figures S1 and S2 (no other pretreatment of the samples was made). The linear response was obtained at $\gamma=1\%$, and all temperatures were 25°C. As long as clay platelets are added, the complex viscosity is quite linear ($|\eta^*| \sim \omega^{-1}$), which indicates gel-like behavior. Towards fastest frequencies ($\omega>50$ rad/s) the inertia of the Couette SANS shear cell comes into play, and the apparent viscosity increases independently of the sample. The inertia is more prominent in the loss tangent at already $\omega>10$ rad/s, except for the very low viscosities of the water/MMT500 system. At rather low frequencies ($\omega<0.1$ rad/s) the Deborah number [41, 42] comes into play, and the Brownian motion of the clay particles in water cause misalignment. That is the reason, why we chose the reference frequency $\omega=1$ rad/s where the signal is strong, and the effect of the tool inertia contribution is weak. The microemulsion/clay systems display a clear trend of decreasing viscosity with increasing platelet diameter (0.1 rad/s $<\omega<50$ rad/s), again with domination of Brownian motions at around $\omega<0.1$ rad/s and influence of the tools inertia at $\omega>50$ rad/s (10 rad/s). The pure microemulsion shows a Newtonian fluid behavior at frequencies $\omega > 10^{-2}$ rad/s. At lowest frequencies, rather elastic behavior is observed.

While the oscillatory experiments proved to be linear and quite reliable at $\omega=1$ rad/s, the steady shear experiments (Figure S3) show instabilities that are related to periodic ordering and disordering of the clay particles [43-46], while the oscillatory experiments give clear hints to stable particle ordering that was initiated from the highest applied frequencies. Only when going to lowest frequencies the Brownian motion destroyed the alignement. Some steady shear experiments achieved a higher degree of ordering (alignment of domains) by applying the shear for longer times of at least 500s, which is much longer as in the case of oscillatory shear.

We observed the stability of clay dispersions by optical inspection, and by combustion experiments (after 7 days). The systems with WXFN140 already showed sedimentation after 30mins, but all shear experiments kept the dispersion stable and optically homogenous throughout the rheological measurements. All other aqueous and microemulsion systems stayed optically unchanged for at least 7 days, the degree of dispersion of which was also confirmed by combustion experiments. The combustion experiments on the crude oil samples showed stable dispersion of ca. 1.1 to 1.2%wt clay in the Colorado crude oil over 8 days, and a decaying dispersion starting from 0.6 to 0.7%wt with a half life of ca. 4 to 5 days in the Pennsylvania crude oil.

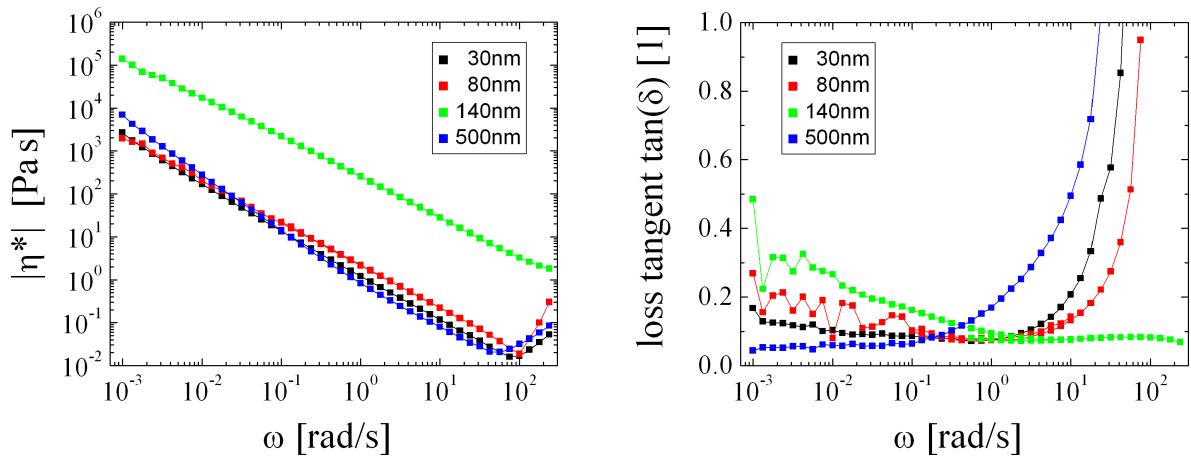

**Figure S1**: The complex viscosity and the loss tangent of frequency sweeps of the aqueous systems with the 4 clays: LRD30, LOG80, WXFN140, MMT500 at 1%vol particle concentration, $\gamma$ = 1% and T = 25°C. All samples appear quite solid like for $\omega$<50 rad/s, for $\omega$<0.1 rad/s the samples turn over to be Brownian motion dominated. For $\omega$>50 rad/s clearly the inertia of the Couette SANS-shear cell dominates, even more visible for tan($\delta$) at around $\omega$ > 10 rad/s.

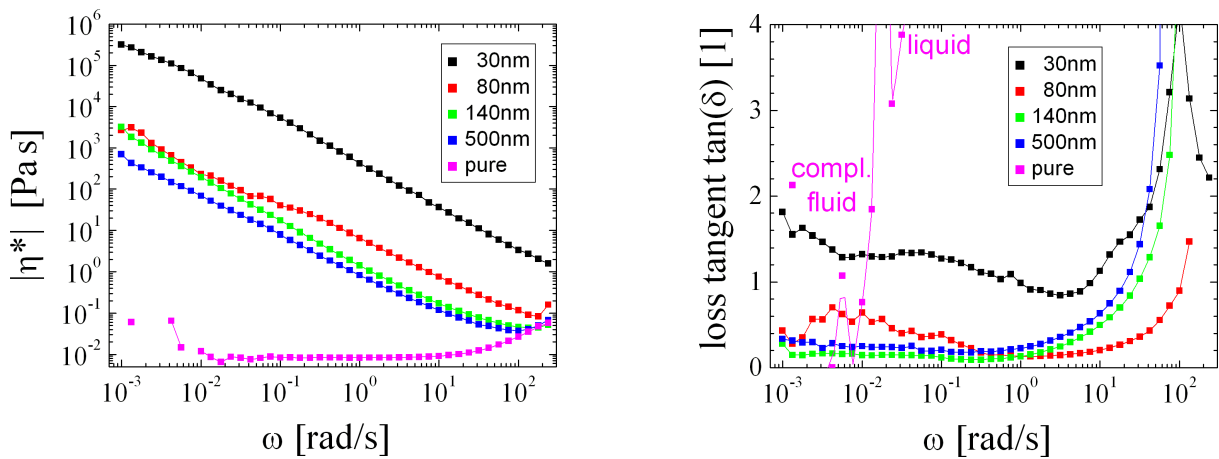

**Figure S2**: The complex viscosity and the loss tangent of frequency sweeps of the microemulsion systems with the 4 clays: LRD30, LOG80, WXFN140, MMT500 at 1%vol particle concentration and without clay, $\gamma$ = 1% and T = 25°C. The clay samples appear quite solid like for $\omega$<50 rad/s, for $\omega$<0.1 rad/s the samples turn over to be Brownian motion dominated. For $\omega$>50 rad/s the inertia of the Couette SANS-shear cell comes into play, even more visible for tan($\delta$) at around $\omega$ > 10 rad/s. The pure microemulsion displays a complex fluid behavior with considerable elasticity at $\omega$<0.01 rad/s, and an ideal fluid behavior at $\omega$>0.05 rad/s.

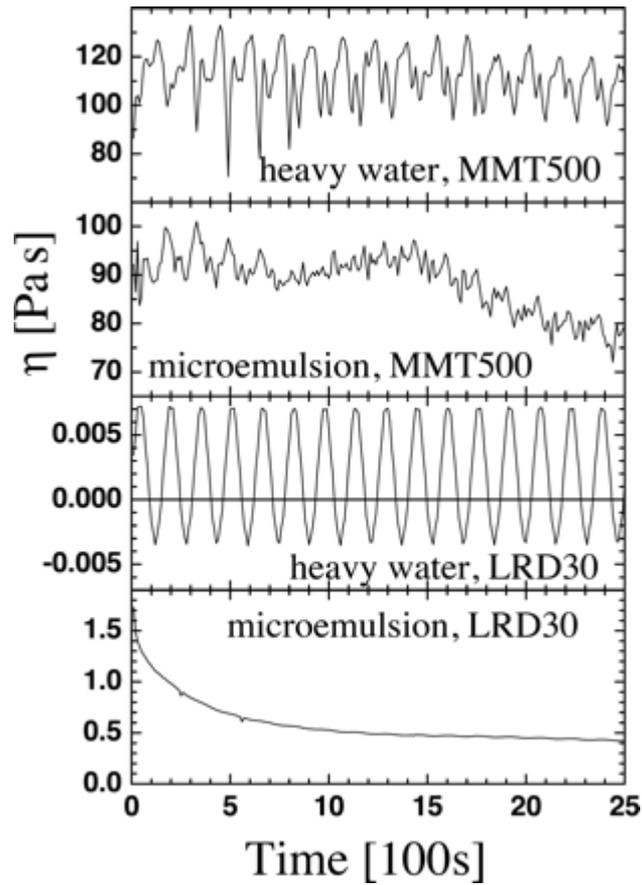

**Figure S3**: Constant shear tests of selected systems during the SANS experiments (viscosity η as a function of time, $\dot{\gamma} = 1s^{-1}$). Quite considerable instabilities are observed in the first three examples, and alignment of domains with time in the last (and second) example.

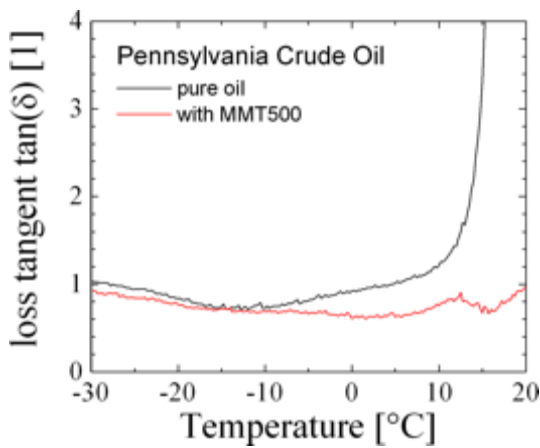 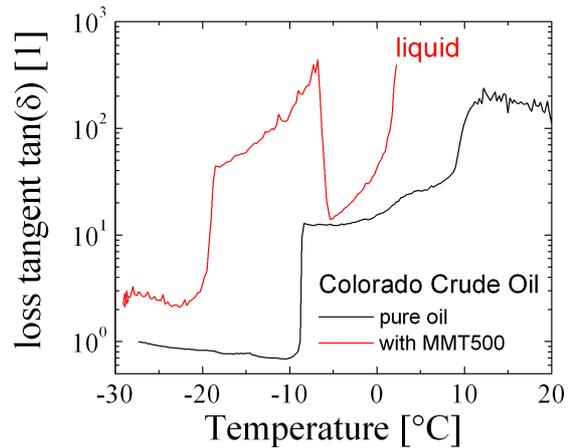

**Figures S4(left) and S5(right)**: The loss tangent of the cooling experiments on crude oils of oscillatory shear (ω=10 rad/s, γ=1%) using cone-plate geometry (plate diameter 5cm). While the Pennsylvania crude oil displays more liquid behavior for the pure oil, the aspects are reversed in the case of the Colorado crude oil.

The missing loss tangent of the oscillatory shear experiment on the crude oils under cooling to ca. -30°C is displayed in Figures S4 and S5. While the Pennsylvania crude oil is more liquid without particles as expected, the Colorado crude oil displays higher elasticity without particles when compared to its mixture with MMT500. However, the complex viscosities showed a clear reduction of the mixtures compared to the corresponding pure oils.

The DSC scans (Figure S6) generally show the produced heat of the flocculation on cooling from 100°C down to 5°C or 10°C. The Pennsylvania crude oil continuously changes over to wax crystallization below 5°C. It does not show a considerable effect by adding the MMT clay. The Colorado crude oil without clay shows a stronger flocculation at around 20°C, which is more evenly distributed for the clay containing sample. Here, the crystallization is well separated at temperatures below 5°C. The signal in between ca. 10°C and 5°C indicates a narrow range of no heat flow and very little heat flow. The latter one might be connected to very high molecular weights.

The $H^1$-NMR scans were performed in dichloromethane (Figure S7). The majority of hydrogens (>90%) is bound in aliphatic fractions (ppm ~ 1 to 2). A smaller fraction is found at around ppm = 2.17 to 4.2. In this fraction the aliphatic neighbors to aromatic rings is found. The whole aromatic fraction is found for ppm = 6.5 to 8.2. Integrals were performed and listed in Table S1. We also tried to precipitate fractions by cooling to -20°C and taking samples from the top. No considerable changes in the integrals were found.

By elementary analysis we determined atomic fractions of the elements C, H, N, and S (see Table S1). Especially the contribution from sulphur could produce an $H^1$-NMR signal in the range of 2.3ppm. From the number fractions, the sulphur does not contribute strongly enough to the $H^1$-NMR signals between 2.17 and 4.2ppm (when looking at the last column of Table S1). So we tend to interpret that this fraction of hydrogens represents the fraction of molecules with aliphatic and aromatic properties. A threshold value of ca. 2% for the $H^1$-NMR signals between 2.17 and 4.2ppm would indicate whether the clay particles would support the lower viscosities at low temperatures or not. We tend to believe that this value is proportional to the number of mediating molecules – similar to surfactants in the microemulsion – and too high fractions would lead to rather molecularly random liquids rather than domain-wise ordered liquids [47].

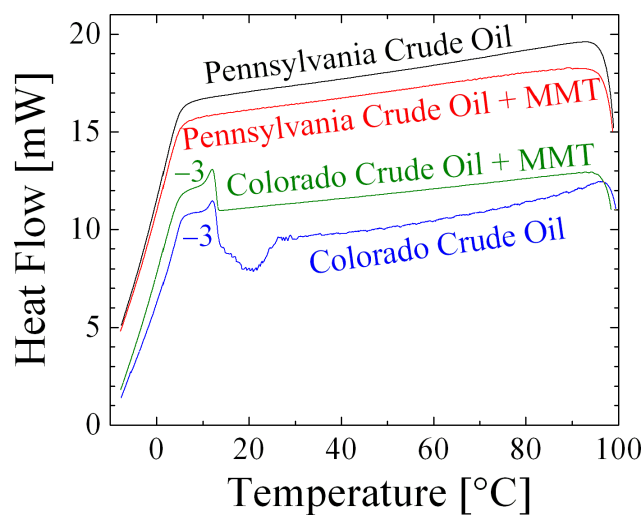

**Figure S6**: Differential Calorimetry Scans on cooling from 100°C to -10°C. The curves of the Colorado crude oil were shifted by -3mW.

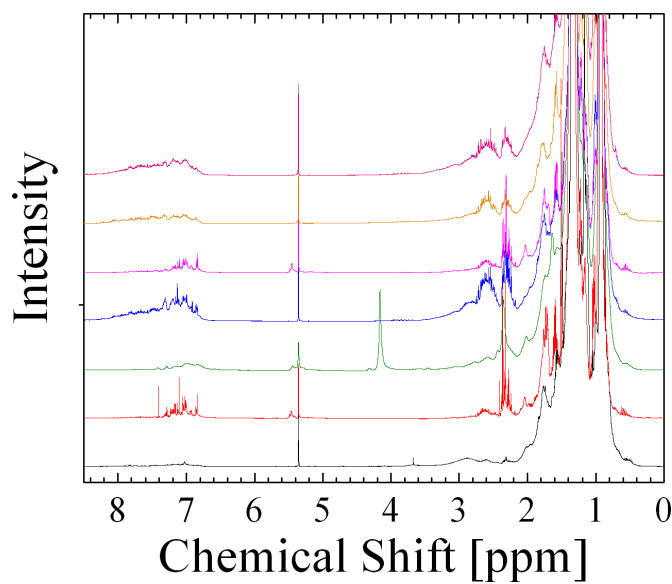

**Figure S7**: $H^1$-NMR measurements of the crude oils in dichloromethane. All spectra were normalized to the integral. The peak at ca. 5.4ppm results from the solvent. Aromatic contributions are found in between 6.5 and 8.2ppm. Aliphatic contributions between 0 and 4.2ppm. We made a cut at 2.17ppm (see text). The oils were (from bottom to top): Colorado crude oil (from main manuscript), Pennsylvania crude oil (from main manuscript), Crudhoe Bay Field, West Texas Intermediate, Guffy Field Pennsylvania, Wattenberg Field, and Gunashli Field crude oils.

**Table S1**: Summaries of H[1]-NMR and elementary analysis. NMR data are number percent. Elementary analysis delivers weight percent. Last column is a guess of number fractions S/H in percent from the elementary analysis.

| Name | 0-2.17ppm | 2.17-4.2ppm | 6.5-8.2ppm | C | H | N | S | S/H |
|---|---|---|---|---|---|---|---|---|
| Colorado Crude Oil | 97.74% | 1.88% | 0.38% | 84.0 ±0.1 | 13.4 ±0.1 | 0.08 ±0.01 | 0.49 ±0.14 | 0.11% |
| Pennsylvania Crude Oil | 96.96% | 1.82% | 1.18% | 76.1 ±1.3 | 12.8 ±0.3 | < 0.08 | <0.19 | 0.02% |
| Crudhoe Bay Field | 95.26% | 3.72% | 0.96% | 82.5 ±0.4 | 13.5 ±0.2 | < 0.08 | 1.46 ±0.02 | 0.34% |
| West Texas Intermediate | 86.63% | 8.32% | 5.03% | 84.7 ±0.5 | 12.6 ±0.1 | 0.08 ±0.01 | 1.80 ±0.01 | 0.45% |
| Guffy Field Pennsylvania | 96.40% | 2.38% | 1.20% | 84.3 ±0.8 | 13.8 ±0.2 | < 0.08 | <0.19 | 0.02% |
| Wattenberg Field | 93.03% | 4.75% | 2.19% | 85.6 ±1.2 | 13.1 ±0.2 | < 0.08 | 0.65 ±0.01 | 0.16% |
| Gunashli Field | 88.84% | 7.20% | 3.88% | 86.0 ±0.2 | 12.2 ±0.1 | 0.18 ±0.01 | 0.32 ±0.03 | 0.08% |

While the physical interpretation of the factor α is given in equation 11, a practical engineering formula [38] for mixtures of water and ethylene glycol was found:

$$\alpha = \exp\left(\phi(1-\phi)\exp\left[a\phi^2\sqrt{T} + \frac{b}{T} - c\right]\right)$$

(S1)

The temperature $T$ is given an absolute Kelvin units, and the constants read $a$ = 33.66 K$^{-1/2}$, $b$ = 986 K, $c$ = 1.5782 for water/ethylene glycol. The original concentrations $\phi$ were molar ratios. However, the order of magnitude of α can be estimated by $\phi$ = 0.5, and $T$ = 300K to be 9. Thus, the order of magnitude of α is explained well.

Apart from that, another detail about α might lie in the interpretation of the number of components. While a microemulsion consists at least of a surfactant and water and oil, the main focus lies on the main components water/oil. Thus the actual 3-component system tends to be interpreted as an apparent two-component system, i.e.

$$\eta_{\text{liq}} = \eta_{\text{oil}}^{\phi_{\text{oil}}} \cdot \eta_{\text{water}}^{\phi_{\text{water}}} \cdot \eta_{\text{surfactant}}^{\phi_{\text{surfactant}}} = \alpha \cdot \eta_{\text{oil}}^{\phi'_{\text{oil}}} \cdot \eta_{\text{water}}^{\phi'_{\text{water}}}$$

(S2)

with $\phi'_{\text{oil}} = \phi_{\text{oil}}/(\phi_{\text{oil}}+\phi_{\text{water}})$, $\phi'_{\text{water}} = 1 - \phi'_{\text{oil}}$ and $\ln(\alpha) = (\phi_{\text{oil}}-\phi_{\text{oil}}/(\phi_{\text{oil}}+\phi_{\text{water}}))\ln(\eta_{\text{oil}}) + (\phi_{\text{water}}-\phi_{\text{water}}/(\phi_{\text{oil}}+\phi_{\text{water}}))\ln(\eta_{\text{water}}) + (\phi_{\text{surfactant}})\ln(\eta_{\text{surfactant}})$. For surfactant concentrations around 20%, and a surfactant viscosity of 1000 times the viscosity of oil and water, one would obtain an apparent α ≈ 4. While in-plane viscosities $\eta_3$ [48] do not differ very much from water and oil, the pure surfactants do have large viscosities. Both formulae (a) and (b) will multiply, i.e. $\alpha_{\text{tot}} = \alpha_{(a)} \alpha_{(b)}$, if both interpretations apply.